\documentclass[10pt,english,english,nofootinbib,notitlepage,superscriptaddress,aps,prd]{revtex4-1}
\usepackage[colorlinks=true, pdfstartview=FitV, linkcolor=blue, citecolor=red, urlcolor=magenta]{hyperref}
\usepackage{graphicx}
\usepackage{latexsym}
\usepackage{amsmath}
\usepackage{amsfonts}
\usepackage{amssymb}
\usepackage{verbatim}
\usepackage{dcolumn}
\usepackage{amssymb}
\usepackage{bm}
\usepackage{float}
\usepackage{subfigure}
\usepackage[english]{babel}
\usepackage{tikz-cd}
\newcommand{\be}{\begin{equation}}
\newcommand{\ee}{\end{equation}}
\newcommand{\bea}{\begin{eqnarray}}
\newcommand{\eea}{\end{eqnarray}}

\begin{document}

\newcommand{\JPess}{
\affiliation{Departamento de F\'isica, Universidade Federal da Para\'iba, \\Caixa Postal 5008, 58059-900, Jo\~ao Pessoa, PB, Brazil}
}

\newcommand{\Lavras}{
\affiliation{Departamento de F\'isica, Universidade Federal de Lavras, Caixa Postal 3037, 37200-000 Lavras-MG, Brazil}
}

\newcommand{\Paris}{
\affiliation{Laboratoire de Physiqe Nucléaire et de Hautes Energies (LPNHE),\\
    Universit\'e Pierre et Marie Curie,\\
    Case courrier 200, 4 place Jussieu,   F-75005 Paris, France}
}

\title{Effects of Planck-scale-modified dispersion relations on the thermodynamics of charged black holes}

\author{I. P. Lobo}
\email{iarley.lobo@ufla.br} 
\email{lobofisica@gmail.com}
\Lavras
\author{V. B. Bezerra}
\email{valdir@fisica.ufpb.br}
\JPess
\author{J. P. Morais Gra\c ca}
\email{jpmorais@gmail.com}
\JPess
\author{Luis C. N. Santos}
\email{luis.santos@ufsc.br}
\JPess
\author{M. Ronco}
\email{mronco@lpnhe.in2p3.fr}
\Paris

\begin{abstract}
Considering corrections produced by modified dispersion relations on the equation of state parameter of radiation, we study the induced black hole metric inspired by Kiselev's ansatz, thus defining a deformed Reissner-Nordstr\"{o}m metric. In particular, we consider thermodynamic properties of such a black hole from the combined viewpoints of the modified equation of state parameter and the phenomenological approach to the quantum gravity problem called rainbow gravity.
\end{abstract}


\maketitle


\section{Introduction}

General relativity is one of the most successful theories in physical sciences, that describes with great precision systems from the scale of the solar system up to the Universe in a large scale by the $\Lambda$CDM paradigm, in particular the remarkable detection of gravitational waves by the merging of black holes and neutron stars are some of the most recent achievements of this theory.
\par
However, in certain regimes, like in the vicinity of the big bang singularity or in the deep interior of a black hole, where quantum effects becomes relevant, it is expected that general relativity or even alternative theories of classical gravity must be replaced by a quantum theory of gravity. There are several theoretical approaches to the problem of quantization of the gravitational field, for instance loop quantum gravity \cite{Rovelli:1997yv}, causal dynamical triangulation \cite{Ambjorn:2010rx}, causal sets approach \cite{Henson:2006kf}, string theory \cite{Mukhi:2011zz}, etc. Despite these theoretical efforts, nowadays one of the main focuses on this problem has been the search for observational traces of a quantum spacetime \cite{AmelinoCamelia:2008qg}.
\par
As described in the review \cite{AmelinoCamelia:2008qg}, one of the most fruitful of these approaches has been the search for Lorentz invariance violation or deformation. In particular, a deformation of this symmetry means that, like Galilean transformations can be found from a ``low-velocity'' limit of the Lorentz symmetry that presents an invariant velocity scale given by the speed of light in vacuum, the Lorentz symmetry could correspond to a ``low-energy'' limit of a deeper symmetry principle that involves an invariant energy scale, i.e. the scale of quantum gravity (which is expected to be of the order of the Planck energy) \cite{AmelinoCamelia:2000mn,AmelinoCamelia:2000ge}. In this case, a question emerges naturally: is there an effective spacetime that can manifest these deformed symmetries? There are indeed some candidates that are able to absorb this property, for instance $\kappa$-Minkowski noncommutative spacetime \cite{Lukierski1,Lukierski2}, Finsler geometry \cite{Girelli:2006fw,Amelino-Camelia:2014rga,Vacaru:2010fa} and the spacetime generated by the relative locality and curved momentum space approach \cite{AmelinoCamelia:2011bm}. These proposals describe deformations of a flat spacetime, where the a deformed Lorentz symmetry plays a fundamental role.
\par
An appealing way to connect this approach to actual astrophysical observations consists in promoting such deformed flat metrics to deformed curved ones, in order to manifest the gravitational field degrees of freedom. There exist some proposals that promote the $\kappa$-Poincar\'e algebra to a curved setup (see \cite{Ballesteros:2019hbw} and references therein), some that explore curved Finsler and Hamilton geometries \cite{Lobo:2016xzq,Letizia:2016lew,Barcaroli:2017gvg}, disformal transformations on a metric \cite{Carvalho:2015omv}, among others. In this paper we shall focus on the simplest and most fruitful of these proposals, called rainbow gravity (RG)\cite{Magueijo:2002xx}, which is defined from an energy-dependent transformation done on the tetrad fields such that one can describe a modified dispersion relation by a norm calculated from an energy-dependent metric, which we shall detail in the next section.
\par   
Within this framework, the issue of black holes and their thermodynamics has been explored by many authors in several different contexts and background theories of gravity, see for instance \cite{Ling:2005bp,Ling:2005bq,Li:2008gs,Gim:2014ira,Hendi:2016oxk,Hendi:2016vux,Hendi:2015cra,Hendi:2016njy,Lobo:2018fym}. Inspired by the Kiselev solution \cite{Kiselev:2002dx}, the case of a quantum corrected black hole surrounded by a fluid with negative equation of state parameter was analyzed in \cite{Banerjee:2016zcu}, where this parameter assumed constant values for different kinds of fluids. However, as raised in \cite{Santos:2015sva}, the fact that we are dealing with modified dispersion relations (MDR) must imply that we need to consider their effects on thermodynamic quantities (leading to temperature-dependent state parameters), leading to a more subtle approach to the initial singularity issue. So, it would be interesting to consider this setup in an astrophysical scenario, like black hole physics. In this paper, we investigate the effect that MDR-compatible thermodynamics has on the thermodynamics of black holes surrounded by radiation in rainbow gravity. Such analysis will lead us to  a proposal for the quantum-corrected charged black hole.
\par
The paper is organized as follows. In Sec. \ref{sec:bh-solution}, we review the definition of rainbow metrics and derive the corrected form of Kiselev's metric from the field equations. In Sec. \ref{sec:EOS}, we perturbatively and numerically analyze the effects that MDRs leave on the equation of state parameter and depict its behavior as a function of the temperature for some specific examples. In Sec. \ref{sec:bh-therm}, we construct the explicit correction of the Reissner-Nordstr\"{o}m metric from combining rainbow gravity, Kiselev's approach, and MDRs, besides that we perturbatively and numerically study the corrected temperature and entropy for each dispersion relation previously considered. In Sec. \ref{sec:alt-appr}, we use an alternative method to study the thermodynamics for a suitable dispersion relation that mimics dust fluid for high energies, and explicitly illustrate this behavior at a trans-Planckian regime. Finally, we conclude in Sec. \ref{sec:conc}.


\section{Rainbow Gravity and Blak Hole solutions}\label{sec:bh-solution}

The possibility of having a deformed Lorentz symmetry driven by a transformation in momentum space was initially raised in the seminal papers \cite{Magueijo:2001cr,Magueijo:2002am}. In this case,  the associated modification of the dispersion relation could be written as:
\begin{equation}\label{mdr1}
m^2=E^2f^2(E/E_P)-p^2g^2(E/E_P),
\end{equation}
where $E$ and $p$ are the energy and norm of the spatial momentum of a fundamental particle and $f(E/E_P)$ and $g(E/E_P)$ are functions of the ratio between $E$ and the invariant Planck energy\footnote{We use a system of units in which the speed of light, Planck's constant and Boltzmann's constant are equal to unity, $c=\hbar=k_B=1$.} $E_P=1/\sqrt{G}$, which we shall assume to be of the order of the quantum gravity energy scale. And they obey the limit $(f,g)\rightarrow 1$, when $E/E_P\rightarrow 0$.
\par
This expression can be equivalently achieved thanks to a map, $U$, that acts on the momentum space as
\begin{equation}
U\triangleright (E,p_i)\doteq U[E,p]=(E\, f(E/E_P),p_i\, g(E/E_P)),
\end{equation}
where the dispersion relation now reads
\begin{equation}\label{mdr1}
m^2=\eta^{\mu\nu}U_{\mu}[E,p]U_{\nu}[E,p],
\end{equation}
where $\eta^{\mu\nu}$ are the components of the Minkowski metric in Cartesian coordinates. In order to absorb this feature into an effective norm using a spacetime metric, we rely on the vielbein of the flat metric. We then transform the orthonormal frame as
\begin{equation}\label{e-tilde}
\tilde{e}_A^{\, \, \mu}=(f(E/E_P)e_{0}^{\, \, \mu},g(E/E_P)e_{I}^{\, \, \mu}),
\end{equation}
which implies that we can rewrite the expression (\ref{mdr1}) as
\begin{equation}
m^2=\eta^{AB}\tilde{e}_{A}^{\, \, \mu}\tilde{e}_{B}^{\, \, \nu}\, P_{\mu}P_{\nu},
\end{equation}
where $P_{\mu}\doteq(E,p_i)$. Until this point, this approach is being applied to the flat spacetime, but its real power relies on applying this transformation on curved vielbeins. This way, it becomes possible to construct an effective curved inverse metric given by $\tilde{g}^{\mu\nu}=\eta^{AB}\tilde{e}_{A}^{\, \, \mu}\tilde{e}_{B}^{\, \, \nu}$, where now we are transforming general curved tetrads according to the prescription (\ref{e-tilde}). Inverting $\tilde{e}_{A}^{\, \, \mu}$, we find an energy-dependent metric 
\begin{equation}\label{r-metric1}
g_{\mu\nu}(E/E_P)=\eta_{AB}\tilde{e}^{A}_{\, \, \mu}\tilde{e}^{B}_{\, \, \nu},
\end{equation}
where $\tilde{e}^{A}_{\, \, \mu}=\left(f^{-1}(E/E_P)e^0_{\, \, \mu},g^{-1}(E/E_P)e^I_{\, \, \nu}\right)$. This is the metric that would be probed by a particle in the intermediate regime in which it is possible to just deform the classical Riemannian geometry by functions that depend on the quantum gravity energy scale, such that when the energy scale of the particles that we are analyzing or the length scale of a given phenomena cannot furnish cumulative effects that might amplify these departures from Riemannian geometry, we recover the usual results from general relativity. The scenario with an emergent metric that depends on the energy (momentum) of the particle itself that travels in such background, originated from quantum corrections, has appeared in different approaches, for instance in the context of analog gravity \cite{Weinfurtner:2008if} and quantization of gravitational degrees of freedom in the Born-Oppenheimer approximation \cite{Assaniousssi:2014ota}.
\par
In order to find the explicit form of the metric probed by these particles for a given spacetime symmetry and matter configuration, we follow the standard approach of finding solutions of the Einstein field equations, where the metric is given by Eq.(\ref{r-metric1}). In spherical coordinates, the static spherically symmetric rainbow metric reads
\begin{equation}\label{r-metric2}
ds^2=-\frac{A(r)}{f^2(\epsilon)}dt^2+\frac{dr^2}{g^2(\epsilon)A(r)}+\frac{r^2}{g^2(\epsilon)}d\Omega^2,
\end{equation}
where we defined the dimensionless variable $\epsilon\doteq E/E_P$,  $d\Omega^2=d\theta^2+\sin^2(\theta)\, d\phi^2$ is the line element of the sphere $S^2$ and the energy is independent of the radial coordinate.
\par
Following Kiselev's approach \cite{Kiselev:2002dx}, after a macroscopic isotropic average, a fluid with energy density $\rho(r)$, pressure $P(r)$ and equation of state $P=\omega\rho$ has an energy-momentum tensor of the form
\begin{eqnarray}
T^t{}_t=T^r{}_r=-\rho(r),\\
T^{\theta}{}_{\theta}=T^{\phi}{}_{\phi}=\frac{1}{2}(3\omega+1)\rho(r).
\end{eqnarray}

Assuming Einstein's field equations $G_{\mu\nu}=\kappa\, T_{\mu\nu}$ ($\kappa=8\pi G$ is the coupling constant) for the rainbow metric (\ref{r-metric2}), a straightforward manipulation implies in the solutions
\begin{eqnarray}\label{metric}
A(r)=1-\frac{2GM}{r}+\frac{c}{r^{3\omega+1}},\label{metric1}\label{metric}\\
\rho(r)=g^2(\epsilon)\frac{c}{\kappa}\frac{3\omega}{r^{3(\omega+1)}}.
\end{eqnarray}

From the expression of the energy density, to preserve its positivity one must have $c\, \omega\geq0$. Besides that there is a factor of $g(\epsilon)$ that corrects the expression found in \cite{Kiselev:2002dx}. Originally, this metric was derived for the analysis of a black hole solution surrounded by exotic fluids, which is described by a negative $\omega$, for instance a quintessence field with $\omega=-2/3$, a cosmological constant $\omega=-1$ (which gives the Schwarzschild-de Sitter solution) or a phantom field with $\omega=-4/3$ \cite{Vikman:2004dc,Caldwell:1999ew}. But this metric can also describe more general kinds of fluids, for instance, if we assume that this spacetime presents an electromagnetic field as matter content, i.e., if $\omega=1/3$, we recover the usual Reisser-Nordstr\"{o}m solution, where $c=Q^2$ is the square of the black hole's electric charge.


\section{Deformed equation of state parameter}\label{sec:EOS}

Modified dispersion relations induce modified equations of state. This issue has been explored in different contexts, for instance in \cite{Santos:2015sva,Nozari:2006yia,Husain:2013zda,Alexander:2001dr}, where the simple nontrivial rule for counting states with given energy leads to unexpected effects, like dimensional reduction or a description of inflation due to modification of the equation of state parameter of radiation, without the need of an inflaton field.
\par
In fact, the number of states with momentum values between $p$ and $p+dp$ in a volume $V$ is given by \cite{huang}
\begin{equation}
N(p)dp=\frac{V}{(2\pi)^3}4\pi p^2 dp,
\end{equation}
and the relation between the momentum and the energy is given now by the deformed expression (\ref{mdr1}). In this paper we are going to consider the effect of these quantum corrections on the Reisser-Nordstr\"{o}m solution, i.e., we shall assume the massless case $m=0$. We basically treat this issue as the classic problem of the photon gas, but with a modified dispersion relation.
\par
Assuming the degeneracy due to the two polarizations of the photons, from Eq.(\ref{mdr1}) a straightforward calculation implies that
\begin{equation}
2N(p)dp\doteq\tilde{N}(E)dE=\frac{V}{\pi^2}\left(\frac{f}{g}\right)^3\left(1+E\frac{f'}{f}-E\frac{g'}{g}\right)E^2,
\end{equation}
where prime ($'$) denotes differentiation with respect to the energy.
\par
As usual, the equation of state parameter can be found from the average of the energy density and the pressure\footnote{Since we are dealing with bosons, the Bose-Einstein statistical distribution is not modified (see Ref.\cite{Alexander:2001dr}, for instance).} as \cite{Santos:2015sva}
\begin{equation}\label{omega1}
\omega(T)=\frac{P}{\rho}=-T\frac{\int \ln[1-e^{-E/T}]\tilde{N}(E)dE}{\int\frac{E}{\exp [E/T]-1}\tilde{N}(E)dE}.
\end{equation}

In this paper, we shall also consider the case of deformation functions depending on the momentum. In these cases, the equation of state parameter reads
\begin{equation}\label{omega2}
\omega(T)=\frac{P}{\rho}=-T\frac{\int \ln[1-e^{-E(p)/T}]p^2 dp}{\int\frac{E(p)}{\exp [E(p)/T]-1}p^2 dp},
\end{equation}
where $E(p)$ shall be determined by the modified dispersion relation under analysis.
\par
For an undeformed dispersion relation, each of these integrals converge to a function given by a power of the temperature, for instance the energy density is proportional to a fourth power of the temperature. In this case, the equation of state parameter is given exactly by the constant value of $1/3$, which corresponds to the usual value associated to a radiation fluid. As stated above, from Kiselev's solution, by assuming this value in Eq.(\ref{metric1}), we derive the metric of a charged black hole.
\par
We then wonder how these quantum corrections affect the Reissner-Nordstr\"{o}m metric in a way that complements the usual rainbow gravity proposal and which imprints they have on the black hole thermodynamics.


\subsection{Some examples of modified dispersion relations}
In this section we are going to describe three particular cases of MDRs that will guide our investigations and the exact behavior of the equation of state parameter as a function of the temperature.
\subsubsection{First case}\label{1case}
As a first example, we consider a case that was first explored in the literature in the context of black hole thermodynamics \cite{Ling:2005bp} and has been a case study for the community of quantum gravity phenomenology ever since:
\begin{equation}\label{function1}
f(\epsilon)=\sqrt{1-\epsilon^2}, \ \ \ \ g(\epsilon)=1.
\end{equation}

In Fig. \ref{MoDR} we depict the behavior of the this MDR for the massless case from Eq.(\ref{mdr1}) represented by the orange (dashed) curve, and we can see the presence of an upper bound on the energy and momentum. Its correction is given by the inverse of the square of the quantum gravity energy scale: 
\begin{equation}\label{ap-mdr1}
E^2= p^2+\frac{E^4}{E_P^2}.
\end{equation}
\par
Besides that, we can define the functional form of the equation of state parameter as a function of the temperature, given implicitly by integrals given by (\ref{omega1}) by the use of (\ref{function1}). Its behavior is depicted in Fig.(\ref{w1}) by the orange (dashed) line. Although we cannot find an analytic form of $\omega$, we can use (\ref{ap-mdr1}) to derive an approximate expression, given by
\begin{equation}\label{appr-omega1}
\omega(T)\approx\frac{1}{3}+\frac{40}{63}\pi^2\frac{T^2}{T_P^2},
\end{equation}
where $T_P$ is the Planck temperature, defined as $T_P=1/\sqrt{G}=E_P$ in the system of units that we are using. For higher temperatures we can also find an analytic expression of $\omega$ in the trans-Planckian regime ($T/T_P\gg 1$) as $\omega(T)\approx\frac{35}{32}\pi \frac{T}{T_P}$.

\subsubsection{Second case}\label{2case}
Now, we explore a dispersion relation that has been recently found in the context of the deformed Lorentz symmetry given by the linear limit of a deformation of the Poisson brackets of general relativity inspired by loop quantum gravity \cite{Brahma:2016tsq}. In one specific case contemplated in \cite{Brahma:2016tsq} the functions that produce the MDR are
\begin{equation}\label{function3}
f(\epsilon)=1, \ \ \ \ g(\epsilon(p))=\sqrt{\frac{2}{\lambda^2p^2}[\lambda p \sinh(\lambda p)-\cosh(\lambda p)+1]}.
\end{equation}

\begin{figure}
\centering
\subfigure[ref1][\, Energy versus momentum.]{\includegraphics[width=7cm,height=4.7cm]{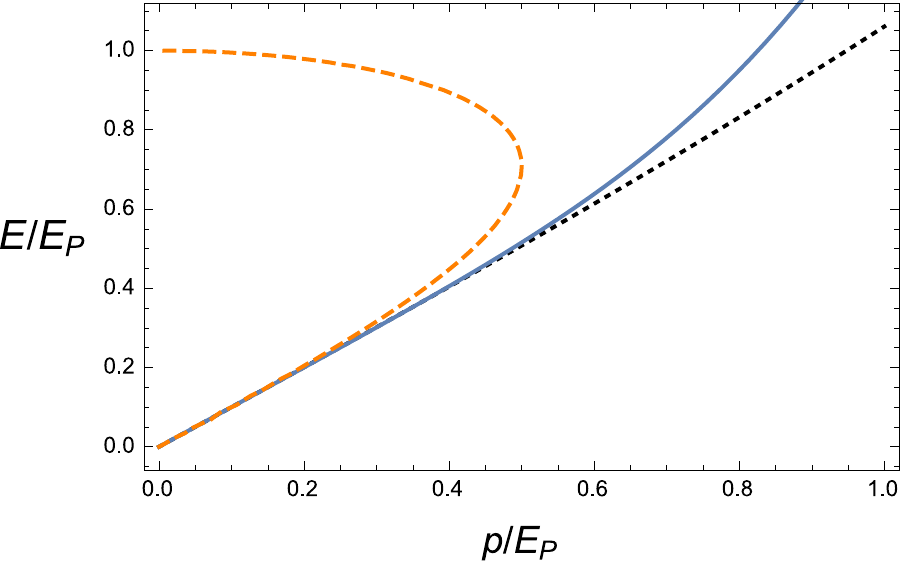}\label{MoDR}}
\qquad
\subfigure[ref2][\, Equation of state parameter versus temperature.]{\includegraphics[width=7cm,height=4.7cm]{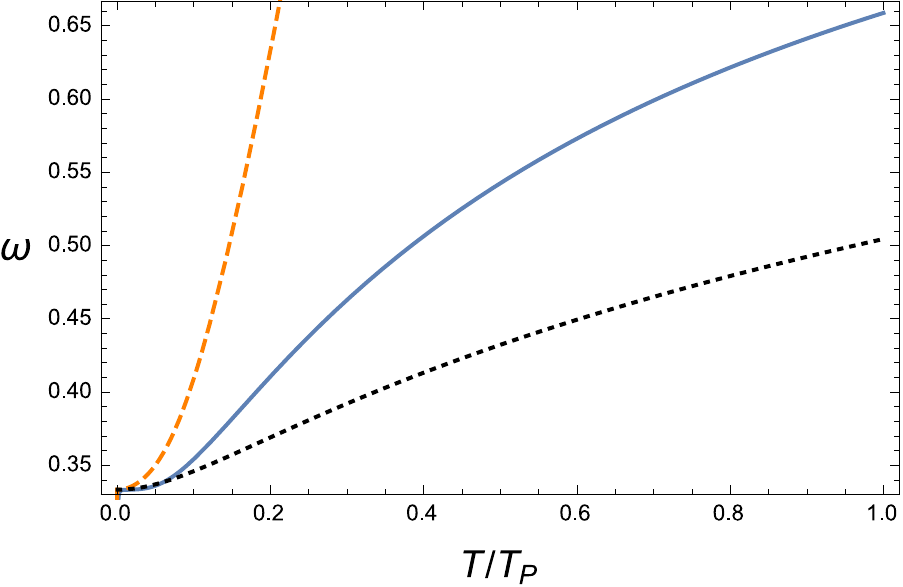}\label{w1}}
\caption{Orange (dashed) line corresponds to $(f,g)=(\sqrt{1-\epsilon^2},1)$, blue (solid) line corresponds to $(f,g)=(1,\sqrt{1+(\lambda p)^4})$, and black (dotted) line corresponds to $(f,g)=\left(1,\sqrt{\frac{2}{\lambda^2p^2}[\lambda p \sinh(\lambda p)-\cosh(\lambda p)+1]}\right)$.}
\end{figure}

This MDR is depicted in Fig. \ref{MoDR} by the black (dotted) line. In this case, there is no energy-momentum upper bound, and its first order approximation is given by
\begin{equation}
E^2\approx p^2+\frac{\lambda^2}{4}p^4.
\end{equation}

As before, the behavior of $\omega$ as a function of the temperature can be found from (\ref{omega1}) and (\ref{function3}) and is depicted in Fig.(\ref{w1}) as the black (dotted) line. In this case, the approximate expression of the equation of state parameter takes the form
\begin{equation}\label{appr-omega2}
\omega(T)\approx\frac{1}{3}+\frac{10}{63}\pi^2\frac{T^2}{T_P^2}.
\end{equation}

This second example consists in an unbounded $\omega$, which surpasses the $+1$ value at a finite temperature around $11\, T_P$.

\subsubsection{Third case}\label{3case}
Our second case of study consists in the following MDR
\begin{equation}\label{function2}
f(\epsilon)=1, \ \ \ \ g(\epsilon(p))=\sqrt{1+(\lambda p)^4},
\end{equation}
that can describe some properties presented in approaches to quantum gravity such as dimensional reduction from causal dynamical triangulation, Horava-Lifshitz gravity, among others Ref.\cite{Ambjorn:2005db,Horava:2009if,Litim:2003vp,Modesto:2008jz,Amelino-Camelia:2013tla}. Here $\lambda$ is simply given by inverse of the Planck energy $\lambda=E_P^{-1}$. The behavior of the MDR for massless particles is depicted in Fig. \ref{MoDR} by the blue (solid) line, where obviously there is no bound in the energy and momentum, in opposition to the first case. It is also given by the expression
\begin{equation}
E^2= p^2+\lambda^4 p^6.
\end{equation}

As before, using Eqs.(\ref{omega1}) and (\ref{function2}), we depict the full behavior of the equation of state parameter in Fig.(\ref{w1}). As also demonstrated in \cite{Santos:2015sva}, the fluid described by this dispersion relation transits from radiation, for low temperatures, with $\omega\approx1/3$ to asymptotically behave like stiff matter for high temperatures, $\omega\approx 1$.
\par
For low temperatures, the first term in the expansion of the equation of state parameter is given by a fourth order power of $\lambda=T_P^{-1}$ :
\begin{equation}\label{appr-omega3}
\omega(T)\approx\frac{1}{3}+\frac{16}{3}\pi^4\frac{T^4}{T_P^4}\, .
\end{equation}


\section{Black Hole Thermodynamics}\label{sec:bh-therm}
We have just verified that the assumption of modified dispersion relations generate a temperature-dependent equation of state parameter for a massless fluid, which can behave in very different ways depending on the model under consideration. Usually, the effect of such deformations is considered on the thermodynamics of black holes in the form of the rainbow functions that corrects the metric, independently from the nature of the undeformed black hole. For instance, a correction to the static charged black hole would be manifest by the procedure that we described in Sec. \ref{sec:bh-solution}, by Eq.(\ref{r-metric2}), where $A(r)$ is the usual Reissner-Nordstr\"{o}m solution.
\par
On the other hand, at the level of the classical gravity, we also demonstrated in Sec \ref{sec:bh-solution} that it is possible to recover the Reissner-Nordstr\"{o}m solution from the Kiselev one, if we assume that the Kiselev fluid has equation of state parameter $\omega=1/3$, which corresponds to the usual radiation fluid.
\par
In this section, we wonder what kind of spacetime we are led to if we consider the effects that MDRs have on the equation of state parameter into the above construction. Specifically, we will construct a quantum-corrected Reissner-Nordstr\"{o}m solution, in a way that complements the usual rainbow gravity approach, from Kiselev's prescription. From this solution, we are going to explicitly write down corrections to the temperature and entropy of the charged black hole, distinguishing the contributions due to rainbow gravity from those coming from a modified dispersion relation.
\par
In order to do this, we shall rewrite the Kiselev solution in a way that enables us to perform a dimensionally coherent analysis. Bearing this in mind, we write the quantum-corrected Kiselev metric as 
\begin{equation}\label{new-metric}
A(r)=1-\frac{2GM}{r}+\left(\frac{Q}{r}\right)^{3\omega(T)+1},
\end{equation}
where $\omega(T)$ is a function of the temperature, given by Eqs.(\ref{omega1}) or (\ref{omega2}), depending on the kind of MDR under consideration. This approach guarantees that the charge $Q$ will have a fixed dimension of length, and will not vary with the temperature. Notice that when $\omega=1/3$, we recover the usual contribution $Q^2/r^2$ to the metric. We summarize our proposal with the diagram (\ref{diagram}), where the map $\psi\circ \phi^{-1}$ is the transformation from the the Reissner-Nordstr\"{o}m solution to the modified solution, driven by the use of a modified dispersion relation on Kiselev's metric.

\begin{equation}\label{diagram}
\begin{tikzcd}[sep=3cm]
& \text{{\it Modified Reissner-Nordstr\"{o}m}}\\
\text{{\it Kiselev}} \arrow[r,"\scalebox{1.5}{$\phi$}", "\text{Dispersion Relation } (m=0)" '] \arrow[ur, swap, "\scalebox{1.5}{$\psi$}", "\text{Modified Dispersion Relation } (m=0)" '] & \text{{\it Reissner-Nordstr\"{o}m}} \arrow[u, swap, "\scalebox{1.5}{$\psi\circ \phi^{-1}$}"]
\end{tikzcd}
\end{equation}
\par
Assuming this approach, we expect to measure some differences regarding the usual procedure followed in rainbow gravity. In fact, usually the Hawking temperature is corrected by a factor of $g(\epsilon)/f(\epsilon)$ \cite{Ling:2005bp}:
\begin{equation}\label{h-temp0}
T_H=-\frac{1}{4\pi}\lim_{r\rightarrow r_+}\sqrt{-\frac{g^{rr}}{g^{tt}}}\frac{1}{g^{tt}}\frac{d}{dr}g^{tt}=\frac{1}{4\pi}\frac{g(\epsilon)}{f(\epsilon)}\left(\frac{2GM}{r_+^2}-(1+3\omega)\frac{Q^{3\omega+1}}{r_+^{3\omega+2}}\right).
\end{equation}
\par
In (\ref{new-metric}), $T$ is the temperature of the radiation fluid that surrounds the black hole. In an equilibrium configuration (which is the one that we assume) it is reasonable to identify this temperature with the one of the fluid of photons emitted by the Hawking radiation emission process. Thus we have, in fact that $\omega=\omega(T_H)$, therefore implying that Eq.(\ref{h-temp0}) is an expression that implicitly defines the Hawking temperature.
\par
As a matter of fact, this procedure presents some similarities with some of the most appealing approaches that demonstrated corrections on the black hole thermodynamics driven by modified dispersion relations \cite{AmelinoCamelia:2005ik} (besides generalized uncertainty principles). In that case, a generalized Stephan-Boltzmann law is calculated considering approximated MDRs, similarly to our calculations of the energy density in the denominator of the equation of state parameter (\ref{omega1}) or (\ref{omega2}). And the temperature of this radiation fluid is also identified with the temperature assigned to the back hole. Based on these arguments, we aim to investigate further corrections of the black hole thermodynamics in our proposed modified Reissner-Nordstr\"{o}m context.
\par
Following the procedure of \cite{Adler:2001vs}, in the vicinity of the black hole, there is an uncertainty in the position of a particle of the order of the Schwarzschild radius $\Delta x=2GM$, and from Heisenberg's uncertainty principle, this can be translated to an uncertainty in the momentum $\Delta x \Delta p\sim 1$, \cite{Adler:2001vs,Gim:2014ira}. We identify this as the momentum of the photon emitted by the Hawking radiation:
\begin{equation}\label{unc}
p=\Delta p\sim \frac{1}{2GM}.
\end{equation}

If we place this value of the momentum into an MDR, it gives us a dependence of the energy of the emitted photon with the mass of the black hole itself. This eliminates the energy dependence of (\ref{h-temp0}), and allows us to study Planck-scale departures induced by a rainbow metric without introducing extra degrees of freedom.
\par
Now, we shall quantify how the assumption of the MDRs chosen in the last section will modify some usual thermodynamic expressions.


\subsection{First case}
As previously stated, our corrected metric has some novelties (due to a modified equation of state parameter) when compared to the the rainbow metric just corrected by the rainbow functions. In the latter case, one usually fixes $\omega=1/3$, while now we allow it to vary with the temperature which, for instance, implies in a temperature dependent location of the horizon, which as a second order approximation is a solution of the transcendental equation
\begin{equation}\label{horizon1}
r_+=Q\exp\left[-\frac{(r_+-\rho_{+})(r_+-\rho_{-})}{\alpha Q^2 T^2/T_P^2}\right],
\end{equation}
where $\rho_{\pm}\doteq GM\pm\sqrt{G^2M^2-Q^2}$ are the locations of the usual Reissner-Nordstr\"{o}m horizons, and $\alpha\doteq \frac{40}{21}\pi^2$ is a dimensionless parameter, whose numerical value will depend on the quadratic corrections at the Planck scale for different MDRs.
\par
From Sec. \ref{1case}, we can make the substitution (\ref{unc}) into the dispersion relation (\ref{mdr1}) by using the rainbow functions (\ref{function1}). Solving the MDR for the energy, we find
\begin{equation}
E=\frac{E_P}{\sqrt{2}}\sqrt{1-\sqrt{1-\frac{1}{G^2M^2E_P^2}}},
\end{equation}
such that $E\approx 1/(2GM)$, when $E_P\rightarrow \infty$, as expected. Solving Eq.(\ref{new-metric}) for $GM$ and substituting it into the temperature, we find the implicit function
\begin{equation}\label{htemp1}
T_H=\frac{r_+^{-1}-3\omega(T)\, Q^{3\omega(T)+1}\, r_{+}^{-2-3\omega(T)}}{2\sqrt{2}\pi\sqrt{1+\sqrt{1-\frac{4r_{+}^{6\omega(T)}}{E_P^2\left[Q^{1+3\omega(T)}+r_+^{1+3\omega(T)}\right]^2}}}},
\end{equation}
where $\omega(T)$ is given by the ratio of integrals (\ref{omega1}), and the integration is done for $E\in[0,E_P)$, since $E_P$ plays the role of an upper bound for the energy for this specific MDR, as can be seen in (\ref{function1}).
\par
\begin{figure}
\centering
\subfigure[ref1][\, $E^2(1-E^2/E_P^2)=p^2$]{\includegraphics[scale=0.67]{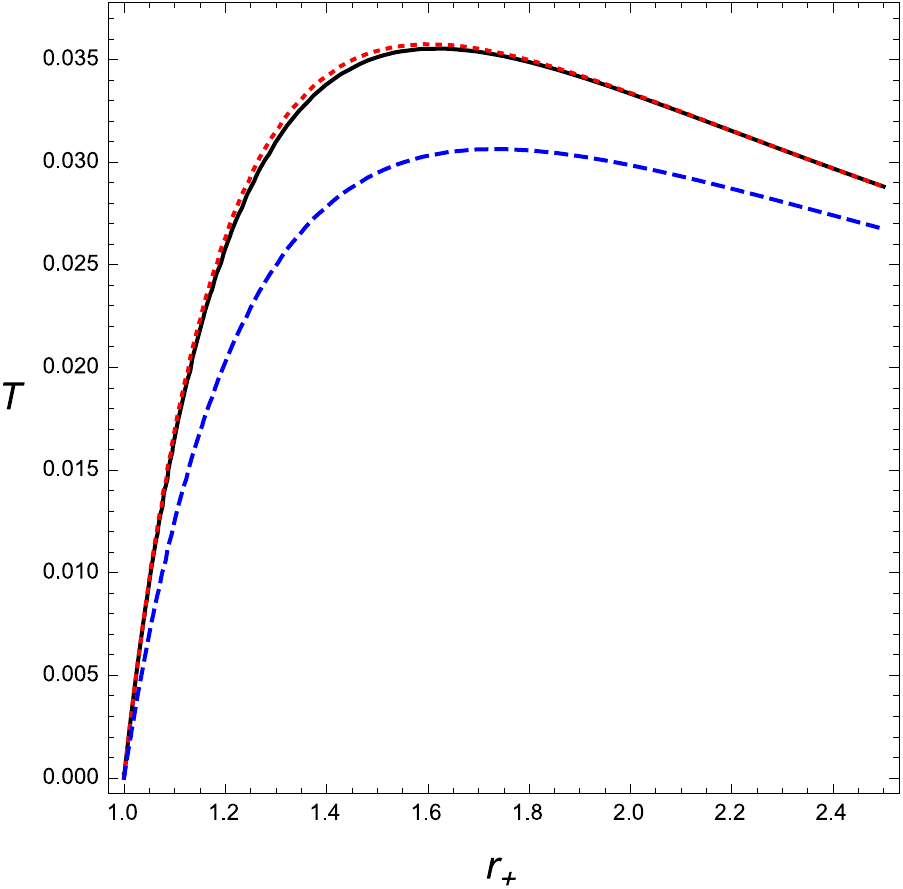}\label{temp1}}
\qquad
\subfigure[ref2][\, $E^2=\frac{2}{\lambda^2}\left(\lambda p \sinh(\lambda p)-\cosh(\lambda p)+1\right)$]{\includegraphics[scale=0.67]{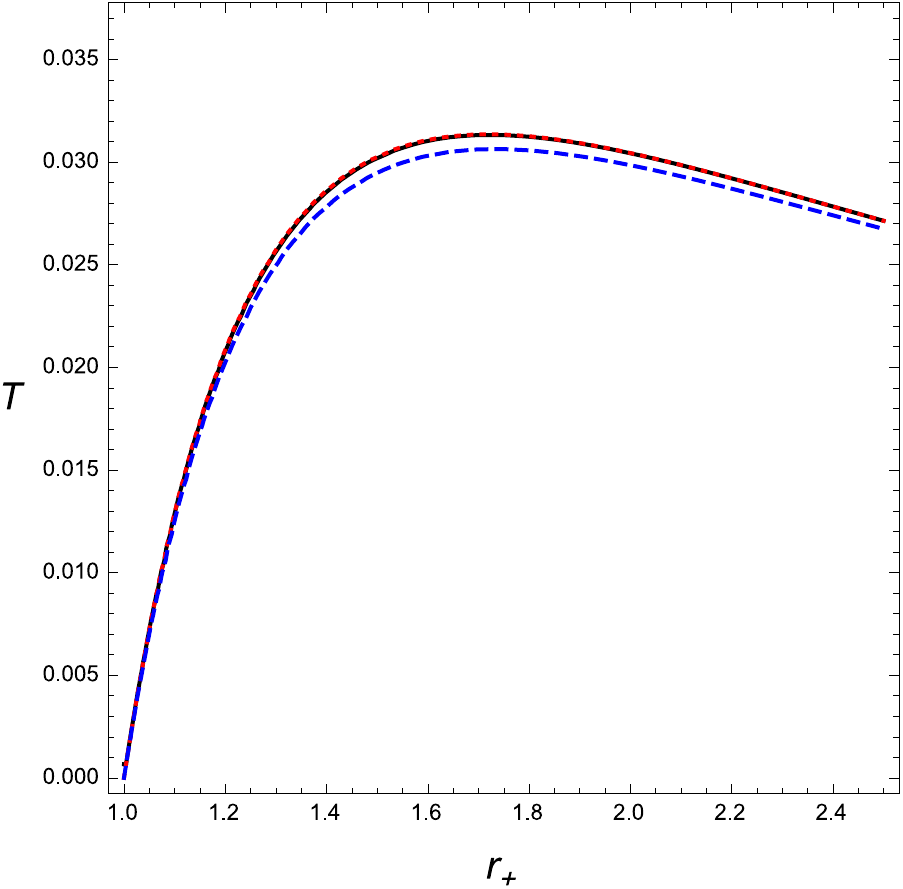}\label{temp2}}
\qquad
\subfigure[ref3][\, $E^2=p^2\left(1+(\lambda p)^4\right)$]{\includegraphics[scale=0.67]{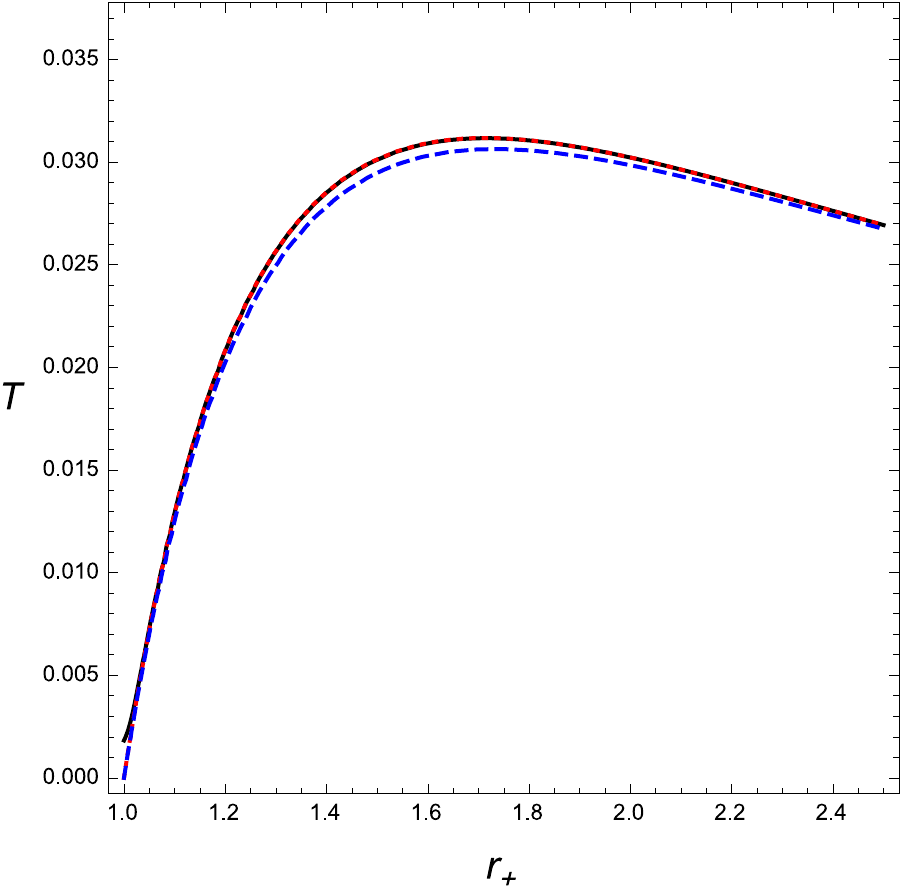}\label{temp3}}
\caption{Implicit plot of the temperature as a function of the horizon radius for $Q=1=E_P$. Blue (dashed) line represents the usual result of general relativity (i.e., when $\omega=1/3$ and $E_P\rightarrow \infty$), the red (dotted) line represents the usual modification of rainbow gravity ($\omega=1/3$), while the black (solid) line describes our novel result.}
\end{figure}

In general, this implicit function cannot be solved analytically; however, it is possible to depict its behavior for some limited range of the horizon radius, as pictured in Fig. \ref{temp1}. The blue (dashed) line corresponds to the case of general relativity, i.e., when $\omega=1/3$ and $f(\epsilon)=1=g(\epsilon)$ (or $E\rightarrow \infty$), and $Q$ equals 1. The completely undeformed case presents a lower bound on the radius given by $r_{\text{lower}}=Q$.
\par
For the usual case in rainbow gravity, one has $\omega=1/3$, but the denominator of Eq.(\ref{htemp1}) is still present. Therefore, there are some extra constraints that must be satisfied by the horizon radius and by the electric charge. In fact, in order for the denominator to be real, we must require that
\begin{equation}
r_+^2\geq 2E_p^{-2}-Q^2\pm 2E_P^{-1}\left(E_P^{-2}-Q^2\right)^{1/2},
\end{equation}
which corresponds to an extra condition, which is an upper bound on the black hole's charge $Q\leq E_P^{-1}$. This is a condition inherited from the MDR. In Fig.(\ref{htemp1}), we considered the extremal case $Q=E_P$, implying that $r_+\geq Q$, which is the usual relativistic result. In our deformed case, we have a complementary condition from the numerator of (\ref{htemp1}) that must remain positive and varies with the temperature.
\par
Although we cannot find an analytic expression for the function $\omega(T)$ from our cases that we are scrutinizing, we can still use first order perturbations in order to quantify departures from general relativity and from the usual rainbow gravity approach.
\par
In fact, using Eq.(\ref{appr-omega1}) (in which the first correction appears quadratic in the inverse of Planck's temperature $T_P$ (energy) and substituting it on our expression for the temperature (\ref{htemp1}), we find the approximate expression 
\begin{equation}\label{new-temp1}
T_{\text{RG+MDR}}\approx T_{\text{RG}}+\frac{5}{168\pi}\frac{Q^2}{T_P^2}\frac{\left(r_+^2-Q^2\right)^2}{r_+^9}\left[\ln \left(\frac{r_+}{Q}\right) -1\right].
\end{equation}
Here, we define $T_{\text{RG+MDR}}$ as the Hawking temperature considering the factor $g(\epsilon)/f(\epsilon)$ due to rainbow gravity (RG) plus the correction from $\omega(T)$ exclusively due to the MDR. And $T_{\text{RG}}$ corresponds just to the correction due to rainbow gravity, for $\omega\equiv 1/3$:
\begin{equation}\label{old-temp1}
T_{\text{RG}}\approx \frac{1}{4\pi}\left(\frac{1}{r_+}-\frac{Q^2}{r_+^3}\right)+\frac{1}{8\pi T_P^2}\frac{r_+^2-Q^2}{r_+\left(r_+^2+Q^2\right)^2}.
\end{equation}

As we can see, for this example of MDR, $T_{\text{RG}}$ presents the ratio of polynomial corrections depending on the horizon radius and the charge. As expected, if $Q=0$, Planck-scale corrections still affect the Schwarzschild black hole, and are of the order $r_+^{-3}$.
\par
Now, analyzing the full contribution $T_{\text{RG+MDR}}$, we verify the presence of a logarithmic correction of the kind $Q^2[\ln(r_+/Q) -1]$, which obviously gets null in the limit $Q\rightarrow 0$. Besides that, the dominant contribution of this correction is of the order ${\cal O}\sim r_+^{-5}\ln(r_+/Q)$, which is highly suppressed when compared to the isolated correction due to the rainbow functions ($r_+^{-3}$). This explains the behavior showed in Fig. \ref{temp1}.

As for the case of the Hawking temperature, our approach manifests new entropy corrections, which can be calculated in a usual way:
\begin{equation}\label{entropy}
S=\int \frac{1}{T}\frac{dM}{dr_+}dr_+.
\end{equation}

Solving the equation $A(r_+)=0$ for the mass, we find $M=M(r_+,Q,\omega(T))$. Using the approximate form of the temperature given by (\ref{new-temp1}) in this expression, we can integrate Eq.(\ref{entropy}) to find
\begin{equation}\label{new-entropy1}
S_{\text{RG+MDR}}=S_{\text{RG}}-\frac{10\pi^2}{21\, GE_P^2}\frac{Q^2}{A^2}\left[A-3\pi Q^2+(A-6\pi Q^2)\ln\left(\frac{A}{4\pi Q^2}\right)\right].
\end{equation}
where
\begin{equation}\label{old-entropy1}
S_{\text{RG}}=\frac{A}{4G}-\frac{\pi}{2GE_p^2}\ln\left(\frac{Q^2}{G}+\frac{A}{4\pi G}\right)-\frac{\pi Q^2}{2GE_P^2\left(A/4\pi+Q^2\right)},
\end{equation}
for $A=4\pi r_+^2$ is the area of the event horizon. As before, the subscript (RG) means usual rainbow gravity correction [calculated using (\ref{old-temp1})], while (RG$+$MDR) means this correction added to those of considering $\omega=\omega(T)$ [calculated from (\ref{new-temp1})].
\par
As it is usually the case in the rainbow gravity literature (for instance \cite{Ling:2005bp}), the function $f(\epsilon)$ and $g(\epsilon)$ allows one to derive logarithmic corrections to the entropy from (\ref{old-entropy1}), which reproduces the kind of behavior found in other approaches to quantum gravity phenomenology, like the use of MDRs and a generalized uncertainty principle that effectively describe results from loop quantum gravity and string theory \cite{AmelinoCamelia:2005ik}. Now, we have extra independent corrections that are only possible from the approach that we propose here for this kind of MDR. As before, in the limiting case of $Q\rightarrow 0$, we recover the results found in \cite{Ling:2005bp}.


\subsection{Second case}

Since the leading order of the dispersion relation defined in Sec. \ref{2case} is also quadratic (in fact it is smaller by a factor of 4), its behavior will be similar to the previous example when we consider just approximate expressions. For instance, the perturbative location of the horizon will be the same as Eq.(\ref{horizon1}), where now $\alpha=10\pi^2/21$.
\par
We follow a procedure similar to the one of the last subsection; however since our rainbow function are a functions directly of the momentum, we do not need to solve an equation for the energy in order to make use of the uncertainty principle. This way, we simply substitute (\ref{unc}) into the dispersion relation (\ref{mdr1}) defined from the set of functions (\ref{function2}).
\par
Solving Eq.(\ref{metric}) for $GM$ and substituting into the temperature (\ref{h-temp0}), we find the implicit function
\begin{equation}
T_H=\frac{E_P r_+^{-2 - 6 \omega}}{2\sqrt{2}\pi}\left[ (Q^{1 + 3 \omega} + r_+^{1 + 3 \omega}) \sqrt{1-\cosh\left[\frac{E_P^{-1}}{r_+ + Q^{1 + 3 \omega} r_+^{-3 \omega} }\right] + \frac{\frac{r_+^{3\omega}}{E_P}\sinh\left[\frac{E_P^{-1}}{r_+ + Q^{1 + 3\omega} r_+^{-3 \omega}}\right]}{Q^{1 + 3 \omega} + r_+^{1 + 3 \omega}}}(r_+^{1 + 3 \omega} - 3 \omega Q^{1 + 3 \omega})\right],
\end{equation}

In this case, the relation between the horizon radius and the charge is modified in our approach because the factor $r_+^{1 + 3 \omega} - 3 \omega Q^{1 + 3 \omega}$ must be positive in order to avoid negative temperatures. A first order approximation for the temperature reads the same expression (\ref{new-temp1}) and (\ref{old-temp1}) of the previous subsection if we map $T_P^2\mapsto 4 T_P^2$. This suppression can be seen in Fig. \ref{temp2} in comparison to Fig. \ref{temp1}, since the values chosen for these plots of the full modified Hawking temperature are such that in the interval $r\in (1,1.5)$, it is completely dominated by this first order term, and in the first case the maximum separation between the result from general relativity and the quantum corrections is numerically of the order $\sim \, 4 \times 10^{-3}$, while for the second case it is of the order $\sim \,  10^{-3}$ i.e., it is four times smaller.
\par
The entropy corrections obey the same rule as the temperature, i.e., it can be found from Eqs.(\ref{new-entropy1}) and (\ref{old-entropy1}), by the map $T_P^2\mapsto 4T_P^2$.


\subsection{Third case}

This third case is similar to the previous subsection in the sense of being unnecessary to solve the MDR for the energy. However, it presents some differences regarding the perturbative expressions of the horizon radius (in this case, not so expressive), the temperature, and the entropy since its leading order comes in a fourth power of the Planck temperature (energy).
\par
As a matter of fact, the definition of the temperature-dependent horizon radius keeps some similarities with the previous case
\begin{equation}\label{horizon3}
r_+=Q\exp\left[-\frac{(r_+-\rho_{+})(r_+-\rho_{-})}{\alpha Q^2 T^4/T_P^4}\right],
\end{equation}
where $\alpha=16\pi^4$. However, the dependence of the temperature comes in a fourth power.
\par
The main departures appear in the temperature and the entropy. In fact, by substituting (\ref{unc}) into the dispersion relation (\ref{mdr1}), using (\ref{function3}), solving Eq.(\ref{metric}) for $GM$ and substituting it into the temperature (\ref{h-temp0}), we find the implicit function
\begin{equation}
T_H=\frac{1}{4\pi}\sqrt{1+\frac{r_+^{12\omega(T)}}{E_P^4\left(Q^{1+3\omega(T)}+r_+^{1+3\omega(T)}\right)}}\left(r_+^{-1}-3\omega(T)\, Q^{3\omega(T)+1}\, r_{+}^{-2-3\omega(T)}\right).
\end{equation}

Also in this case, the positiveness of the temperature imposes a complementary relation between the horizon radius and the charge. Besides that, the ratio of polynomials that correct the temperature in rainbow gravity is different from the previous case, as expected:
\begin{equation}
T_{\text{RG}}\approx \frac{1}{4\pi}\left(\frac{1}{r_+}-\frac{Q^2}{r_+^3}\right)+\frac{r_+}{8\pi T_P^4}\frac{r_+^2-Q^2}{\left(r_+^2+Q^2\right)^4}.
\end{equation}
and the full correction due to considering the MDR also presents important differences in comparison to the previous cases of second order perturbations:
\begin{equation}\label{new-temp3}
T_{\text{RG+MDR}}\approx T_{\text{RG}}+\frac{1}{64\pi}\frac{Q^2}{T_P^4}\frac{\left(r_+^2-Q^2\right)^4}{r_+^{15}}\left[\ln \left(\frac{r_+}{Q}\right) -1\right].
\end{equation}

In fact, this new contribution is highly suppressed for higher values of the horizon, since its dominant terms are of the form $Q^2r^{-7}$, while those from rainbow gravity are of the kind $r^{-5}$. This behavior is explicit in Fig.(\ref{temp3}), where there is practically no difference between the usual rainbow gravity correction and our new contribution for the values considered in our plot. Besides that, the global nature of the Hawking temperature of the undeformed black hole surrounded by radiation and stiff matter are similar (a property that is preserved when considering rainbow corrections), which is an extra explanation for the similarities presented in our graphs.
\par
Also the entropy presents significant differences in comparison to the previous case, in particular the absence of the logarithmic corrections for the pure rainbow gravity case (which is a feature of MDRs that are quadratic on Planck energy). But, we now have a novel result due to the recovery of the logarithmic corrections due to the consideration of a temperature-dependent $\omega$,

\begin{eqnarray}\label{entropy3}
S_{\text{RG+MDR}}=S_{\text{RG}}+\frac{\pi}{G\, E_P^4}\frac{Q^2}{r_+^{10}}\left\{\frac{3}{200} Q^6\left[1+10\ln\left(\frac{r_+}{Q}\right)\right] - \frac{7}{128}Q^4 r_+^2 \left[1+8\ln\left(\frac{r_+}{Q}\right)\right]\right. \nonumber\\
\left.+ \frac{5}{72}Q^2 r_+^4\left[1+6\ln\left(\frac{r_+}{Q}\right)\right] - \frac{1}{32} r_+^6\left[1+4\ln\left(\frac{r_+}{Q}\right)\right]\right\}
\end{eqnarray}
where
\begin{equation}
S_{\text{RG}}=\frac{A}{4G}+\frac{\pi}{6G\, E_P^4}\frac{3r_+^4+3Q^2r_+^2+Q^4}{(Q^2+r_+^2)^3}.
\end{equation}

This new contribution presents striking differences in comparison to the quadratic corrections and presents all possible combinations of products of powers of $Q$ and $r_+$ with dimensions length to the sixth. However, as before this new correction is highly suppressed by a factor of $r_+^{-10}$, as can be seen in (\ref{entropy3}).
\par
In the next section, we shall explore an alternative formulation for dealing with the energy dependence of the metric, that will allow us to analyze a case in which the temperature presents a globally different behavior in comparison to the charged cases of general relativity and rainbow gravity.


\section{Alternative approach}\label{sec:alt-appr}

In this section we shall follow a different approach to this issue that is suitable to situations in which it is not possible to solve the modified dispersion relation for the energy and requires an alternative relation between quantities related to the black hole itself and the energy presented in the rainbow functions. To illustrate this approach, let us work with the following functions:

\begin{equation}\label{functions4}
f(\epsilon)=\frac{e^{E/E_P}-1}{E/E_P},  \ \ \ \  g(\epsilon)=1,
\end{equation}
which in a first order approximation implies the following MDR that presents a linear deformation:

\begin{equation}
E^2\approx p^2-\frac{E^3}{E_P}.
\end{equation}

\begin{figure}[h]
\centering
\subfigure[ref1][\, Energy versus momentum.]{\includegraphics[scale=0.7]{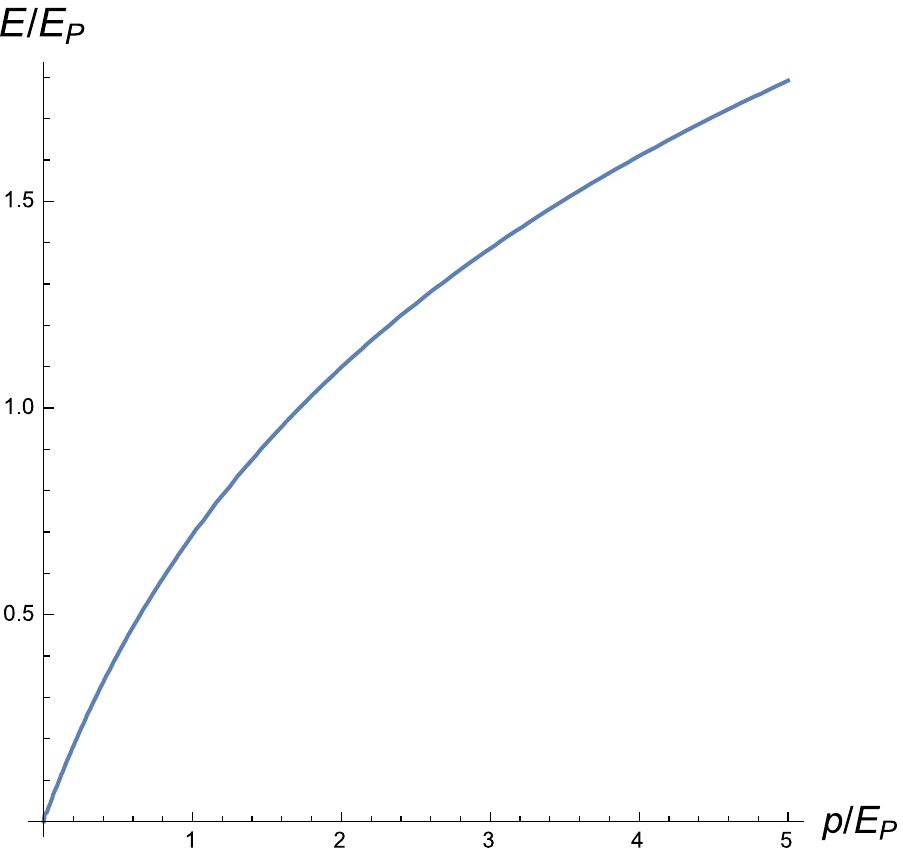}\label{MoDR4}}
\qquad
\subfigure[ref2][\, Equation of state parameter versus temperature. Radiation behaves as dust for high temperatures.]{\includegraphics[scale=0.9]{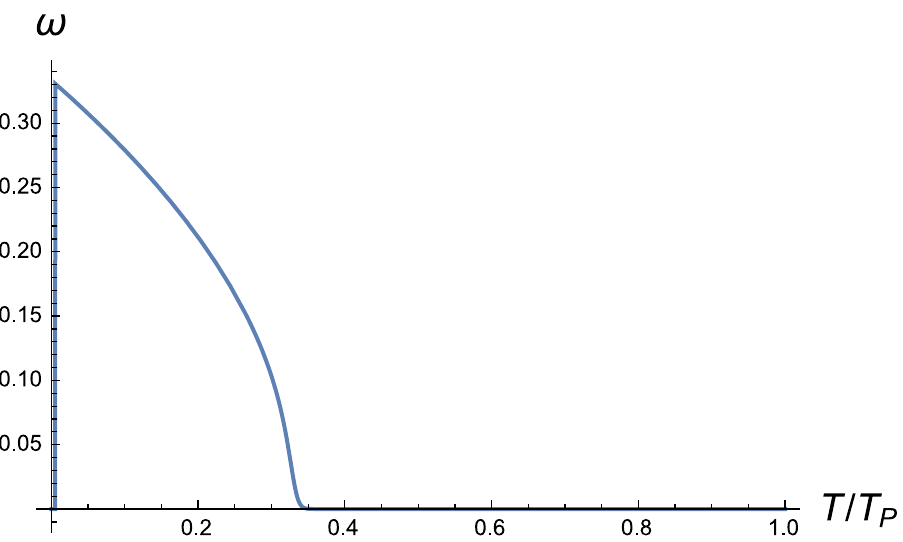}\label{w4}}
\caption{\, $\left(\frac{e^{E/E_P}-1}{E/E_P}\right)^2E^2=p^2$.}
\end{figure}

This has been the case of study of some of the first investigations in quantum gravity phenomenology \cite{AmelinoCamelia:2000mn,AmelinoCamelia:1997gz}, and is depicted in Fig. \ref{MoDR4}. From Eq.(\ref{omega1}), we can define the temperature-dependent equation of state parameter $\omega(T)$, whose behavior is depicted in Fig. \ref{w4}. As can be seen, for low temperatures, the radiation fluid behaves like the usual case, i.e., with $\omega\approx 1/3$, as a matter of fact, its first order functional dependence is the following:
\begin{equation}\label{omega4}
\omega(T)\approx \frac{1}{3}-\frac{60}{\pi^4}\zeta(5)\frac{T}{T_P},
\end{equation}
where we have the presence of the Riemann zeta function $\zeta(s)=\sum_{n=1}^{\infty}\frac{1}{n^s}$, where $\zeta(5)\approx 1.03$. While for Planckian temperatures, this fluid behaves like dust, i.e., with $\omega=0$. This fact had already been noticed in Ref.\cite{Santos:2015sva}, where the authors were concerned with the effect of modified dispersion relations on the equation of state parameter in a cosmological background,

The temperature-dependent horizon can than be straightforwardly read from Eq.(\ref{new-metric}) as
\begin{equation}\label{horizon4}
r_+=Q\exp\left[\frac{(r_+-\rho_{+})(r_+-\rho_{-})}{\frac{180}{\pi^4}\zeta(5) Q^2 T/T_P}\right].
\end{equation}
\par
The Hawking temperature is defined in the usual way, by Eq.(\ref{h-temp0}). However, we now follow an approach different from the last section and let us suppose that we define the rainbow metric from measurements of photons with an average energy $E=\langle E\rangle$ such that we can identify the energy of the photons emitted from Hawking radiation with the black hole temperature, i.e, $T=E$ \cite{Ling:2005bp}.
\par
From this identification, and the definition (\ref{h-temp0}), the Hawking temperature $T$ is implicitly defined by the equation:
\begin{equation}\label{temp4-0}
T=\frac{T}{T_P}\frac{2GMr^{-2}-3\omega(T)Q^{1+3\omega(T)}r^{-2-3\omega(T)}}{4\pi(e^{T/T_P}-1)}.
\end{equation}

\begin{figure}[h]
\includegraphics[scale=0.75]{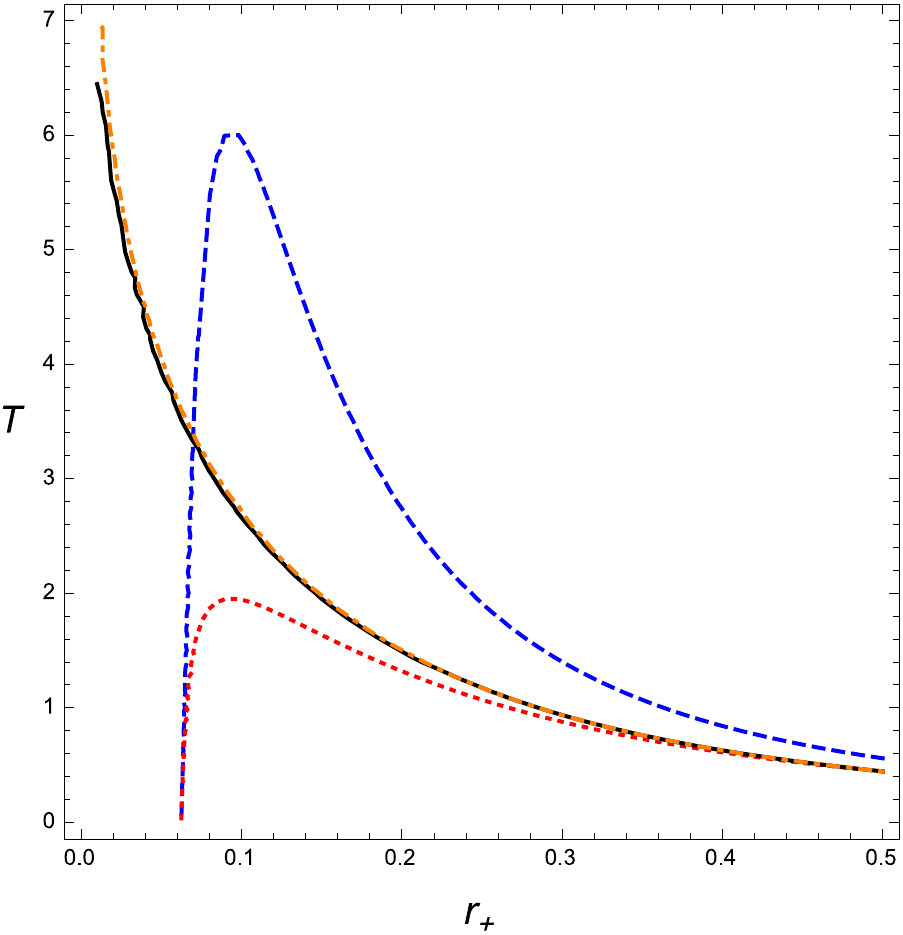}\label{temp4}
\caption{Temperature versus horizon radius for $M=1=E_P$ and $Q=0.5$ and MDR $\left(\frac{e^{E/E_P}-1}{E/E_P}\right)^2E^2=p^2$. For high temperatures, the modified Reissner-Nordstr\"{o}m black hole mimics the Schwarzschild one. Blue (dashed) line represents GR, the red (dotted) line represents RG, the black (solid) line describes RG+MDR, and the orange (dotted-dashed) line describes a RG Schwarzschild black hole.}
\end{figure}

The behavior of this temperature versus the horizon radius is depicted in Fig. \ref{temp4}, where we chose the values $M=1=E_P$ and $Q=0.25$. This way we can see this numerical analysis confirms exactly what we expected from the analysis of the equation of state parameter $\omega$. Since for low temperatures, the black hole behaves like the usual Reissner-Nordstr\"{o}m solution, while in a Planckian regime the black hole temperature starts to simulate Kiselev's solution with $\omega=0$, which corresponds to the Schwarzschild black hole with a redefined mass $\tilde{M}=M-Q/2G$ corrected just by the rainbow gravity factors. 
\par
In that figure, the blue (dashed) line represents the Hawking temperature of the Reissner-Nordstr\"{o}m metric, while the red (dotted) one presents correction just due to rainbow gravity (and our ansatz that relates the temperature and the energy). Now, the black (solid) line is the implicit plot of the temperature (\ref{temp4-0}), where $\omega(T)$ is given by the ratio of integrals given by Eq.(\ref{omega1}) using the functions (\ref{functions4}). In this case, we verify a departure of the Reissner-Nordstr\"{o}m behavior for higher temperatures and a gradual approximation of the Schwarzschild-like case corrected by the rainbow factor. For illustrative reasons, we added an extra curve, the orange (dotted-dashed) one that corresponds this aforementioned rainbow-Schwarzschild case for this choice of parameters.
\par
The corrective factor given by the rainbow functions $g(\epsilon)/f(\epsilon)$ shift the maximum value attained by the temperature without modifying its overall form. This example demonstrates that our corrections have deeper consequences on metrics corrected by modified dispersion relations.

Besides what we already discussed, the presence of corrections linear on the Planck scale, leads to first order equations that differ from the previous approaches. In fact, by using expression (\ref{omega4}) in the implicit Eq. (\ref{temp4-0}) and solving it for $T$, we find a full expression $T_{\text{RG+MDR}}$ that, as before, can be decomposed into a rainbow gravity term $T_{\text{RG}}$, plus our novel contribution:

\begin{equation}
T_{\text{RG+MDR}}\approx T_{\text{RG}} -\frac{45}{4\pi^6 T_P}\zeta(5)Q^2\frac{(r_+^2-Q^2)}{r_+^6}\left[\ln\left(\frac{r_+}{Q}\right)-1\right],
\end{equation}
where
\begin{equation}
T_{\text{RG}}\approx \frac{1}{4\pi}\left(\frac{1}{r_+}-\frac{Q^2}{r_+^3}\right)-\frac{1}{32\pi E_P}\frac{(r_+^2-Q^2)^2}{r_+^6}.
\end{equation}

The entropy corrections are consistently different than the previous cases, for instance the usual rainbow gravity contribution reads

\begin{equation}
S_{\text{RG}}\approx \pi\frac{r_+^2}{G}+\frac{1}{4GT_P}\left(r_++\frac{Q^2}{r_+}\right),
\end{equation}
while our full expression is

\begin{eqnarray}
S_{\text{RG+MDR}}\approx S_{\text{RG}}+\zeta(5)\frac{90}{\pi^4 G E_P}\frac{Q}{r_+}\left\{3Q\left[1+\ln\left(\frac{r_+}{Q}\right)\right]+r\ln\left(\frac{r_+}{Q}\right)\ln\left(\frac{Q-r_+}{Q+r_+}\right)\right.\nonumber\\
\left.+r_+\left[\text{Li}_2\left(\frac{r_+}{Q}\right)-\text{Li}_2\left(-\frac{r_+}{Q}\right)\right]\right\},
\end{eqnarray}
where we verify the nontrivial presence of the polylogarithm function $\text{Li}_s(z)=\sum_{k=1}^{\infty}\frac{z^k}{k^s}$.


\section{Final remarks}\label{sec:conc}

We proposed a phenomenological application of Kiselev's solution of a black hole surrounded by a fluid in the scenario of modified dispersion relations. These solutions basically differ by the equation of state parameters used to model the fluids \cite{Kiselev:2002dx}. As a particular case of study, we considered the radiation case, which is described by the usual $\omega=1/3$ parameter when the massless dispersion relation reads $E^2=p^2$ and furnishes the metric of a static charged black hole (Reissner-Nordstr\"{o}m solution). In a phenomenological scenario that takes into account modified dispersion relations (as typically expected from low energy models of quantum gravity), corrections on $\omega$ must take place. In this case, we are mapped onto a modified Reissner-Nordstro\"{o}m metric that, in principle, could give us information about a quantum regime in a bottom-up approach to quantum gravity.
\par
Complementarily, we also absorb information about a quantum property of spacetime by exploring the, so-called, rainbow metric defined in the seminal paper \cite{Magueijo:2002xx}. This way, we aim to furnish a coherent correction of the quantum corrected charged black hole. In this case, the equation of state parameter depends on the ratio between the temperature of the fluid and the temperature (energy) scale of the modified dispersion relation (which for simplicity we treat as the Planck temperature); therefore, it turns out that the analysis of the corresponding black hole thermodynamics was a natural path to follow.
\par
We then considered four modified dispersion relations (a first analysis with three examples, and a fourth one used for illustrating an alternative formulation), where each of them either comes from a particular theoretical approach to quantum gravity or from particular phenomenological motivations. We were then able to calculate approximative corrections on thermodynamic quantities, such as the Hawking temperature and the black hole entropy, where we isolated those contributions coming exclusively from the rainbow gravity approach from ours. Analytical expressions could not be found for the case of these MDRs, but we depicted the behavior of the equation of state parameter and the Hawking temperature by assuming their implicit definitions given in terms of undefined integrals. In particular, the temperature obeyed the qualitative behavior expected from the asymptotic behavior of the equation of state parameter.
\par
The main results of our analysis are the presence of an uncertainty in the geometrical location of the horizon, depending on the temperature of the fluid of photons used to probe this spacetime. Besides that, we found corrections on expressions for minimal black hole radii (when this is the case), which must have imprints in the calculation of remnants of black hole evaporation, which shall be analyzed in future investigations. We also found logarithmic corrections of the temperature, which were transmitted to the entropy, demonstrating that it is still possible to have this kind of correction (which are common in other approaches to quantum gravity) even when the usual rainbow gravity approach does not furnish them. And the analysis of the temperature indeed demonstrated that for smaller radii (and higher temperatures) a particular black hole can mimic other cases, which could have important consequences, like modifications of the causal structure of these spacetimes, which could be read, for instance, from temperature-dependent Penrose-Carter diagrams.
\par
This approach allows one to investigate quantum corrections for fluids with different equations of state parameters, being necessary just to find where the modified dispersion relation can be important for counting of states in a given momenta interval, which might have imprints on other astrophysical objects studied in quantum gravity phenomenology.


\acknowledgments
The authors would like to thank CNPq (Conselho Nacional de Desenvolvimento Cient\'ifico e Tecnol\'ogico, Brazil) for financial support. This study was financed in part by the Coordena\c{c}\~ao de Aperfei\c{c}oamento de Pessoal de N\'ivel Superior, Brazil (CAPES), Finance Code 001. This article is based upon work from COST Action CA18108, supported by COST (European Cooperation in Science and Technology) .


\end{document}